------------------------------------------------------------------------------

\documentstyle[eqsecnum,aps]{revtex}
\begin{document}

\newcommand{\ds}{\displaystyle}

\title{The Relativistic Levinson Theorem in Two Dimensions}

\author{Shi-Hai Dong\thanks{Electronic address:DONGSH@BEPC4.IHEP.AC.CN}}

\address{Institute of High Energy Physics, P. O. Box 918(4), Beijing 100039,
The People's Republic of China}

\author{Xi-Wen Hou}

\address{Institute of High Energy Physics, 
P. O. Box 918(4), Beijing 100039\\
and Department of Physics, University of Three
Gorges, Yichang 443000, The People's Republic of China}

\author{Zhong-Qi Ma}

\address{China Center for Advanced Science and Technology
(World Laboratory), P. O. Box 8730, Beijing 100080\\
and Institute of High Energy Physics, P. O. Box 918(4),
Beijing 100039, The People's Republic of China}

%\date{}

\maketitle

\vspace{4mm}

\begin{abstract}
In the light of the generalized Sturm-Liouville theorem, the 
Levinson theorem for the Dirac equation in two dimensions is 
established as a relation between the total number $n_{j}$ 
of the bound states and the sum of the phase shifts
$\eta_{j}(\pm M)$ of the scattering states with the angular 
momentum $j$:
$$\eta_{j}(M)+\eta_{j}(-M)~~~~~~~~~~~~~~~~~~~~~~~~~~~~~~~~~~
~~~~~~~~~~~~~~~~~~~~~~~~~~~~~~~~~~~~~~~~~~~~~~~~~~~~~~~~~~~~$$
$$~~~=\left\{\begin{array}{ll}
(n_{j}+1)\pi
&{\rm when~a~half~bound~state~occurs~at}~E=M ~~{\rm and}~~
j=3/2~{\rm or}~-1/2\\
(n_{j}+1)\pi
&{\rm when~a~half~bound~state~occurs~at}~E=-M~~{\rm and}~~
j=1/2~{\rm or}~-3/2\\
n_{j}\pi~&{\rm the~rest~cases} . \end{array} \right.  $$

\noindent
The critical case, where the Dirac equation has a finite 
zero-momentum solution, is analyzed in detail. A zero-momentum
solution is called a half bound state if its wave function
is finite but does not decay fast enough at infinity to be 
square integrable.

\end{abstract}

\vspace{10mm}

\section{INTRODUCTION}

The Levinson theorem [1] is an important theorem in the 
quantum scattering theory, which sets up the relation between 
the number of bound states and the phase shift at zero momentum. 
It has been generalized [2-9] and applied to different fields 
in modern physics [10-16].  Recently, the Levinson theorem in 
two dimensions was studied both in experimental [17] and 
theoretical [18-20] aspects because of the wide interest in 
the lower dimensional field theories.

In this paper we will study the Levinson theorem for the Dirac 
equation in two dimensions:
$$\displaystyle \sum_{\mu=0}^{2}i\gamma^{\mu}
\left(\partial_{\mu}+ieA_{\mu}\right)\psi=M\psi, 
\eqno (1) $$

\noindent
where $M$ is the mass of the particle, and 
$$\gamma^{0}=\sigma_{3},~~~~~
\gamma^{1}=i\sigma_{1},~~~~~\gamma^{2}=i\sigma_{2}. \eqno (2) $$

\noindent
Throughout this paper the natural units $\hbar=c=1$ are employed. 
Discuss the special case where only the zero component of 
$A_{\mu}$ is non-vanishing and cylindrically symmetric:
$$A_{1}=A_{2}=0,~~~~~eA_{0}=V(r). \eqno (3) $$

\noindent
The boundary condition at the origin for the potential $V(r)$ is
necessary for the nice behavior of the wave function
$$\displaystyle \int_{0}^{1}r|V(r)|dr < \infty. \eqno (4) $$

\noindent
For simplicity, we firstly discuss the case where the potential 
$V(r)$ is a cutoff one at a sufficiently large radius $r_{0}$:
$$V(r)=0,~~~~~{\rm when}~~r\geq r_{0}. \eqno (5) $$

\noindent
The general case where the potential $V(r)$ has a tail
at infinity will be discussed in Sec.V.

Introduce a parameter $\lambda$ for the potential $V(r)$:
$$V(r,\lambda)=\lambda V(r) . \eqno (6) $$

\noindent
As $\lambda$ increases from zero to one, the potential
$V(r,\lambda)$ changes from zero to the given potential 
$V(r)$. If $\lambda$ changes its sign, the potential
$V(r,\lambda)$ changes sign, too.

Letting 
$$\psi_{jE}(t,{\bf r},\lambda)=e^{-iEt}r^{-1/2}\left(\begin{array}{c}
f_{jE}(r,\lambda)e^{i(j-1/2)\varphi} \\ 
g_{jE}(r,\lambda)e^{i(j+1/2)\varphi} \end{array} \right), \eqno (7) $$

\noindent
where $j$ denotes the total angular momentum, $j=\pm 1/2$,
$\pm 3/2$, $\ldots$, we obtain the radial equations:
$$\displaystyle {d \over dr}g_{jE}(r,\lambda)+\displaystyle 
{j \over r} g_{jE}(r,\lambda) 
=\left(E-V(r,\lambda)-M\right)f_{jE}(r,\lambda) , $$
$$-\displaystyle {d \over dr}f_{jE}(r,\lambda)+\displaystyle 
{j \over r} f_{jE}(r,\lambda) 
=\left(E-V(r,\lambda)+M\right)g_{jE}(r,\lambda) .  \eqno (8) $$

\noindent
It is easy to see that the solutions with a negative $j$ can be
obtained from those with a positive $j$ by interchanging 
$f_{jE}(r,\lambda) \longleftrightarrow g_{-j-E}(r,-\lambda)$, 
so that in the following we only discuss the solutions with 
a positive $j$. The main results for the case with a negative 
$j$ will be indicated in the text.

The physically admissible solutions are finite, continuous, 
vanishing at the origin, and square integrable:
$$f_{jE}(r,\lambda)=g_{jE}(r,\lambda)=0,~~~~~
{\rm when}~~r\longrightarrow 0, \eqno (9) $$
$$\displaystyle \int_{0}^{\infty} dr\left\{|f_{jE}(r,\lambda)|^{2}
+|g_{jE}(r,\lambda)|^{2}\right\} < \infty. \eqno (10) $$

\noindent
The solutions for $|E|>M$ describe the scattering states, and
those for $|E|\leq M$ describe the bound states. We will solve
Eq.(8) in two regions, $0\leq r < r_{0}$ and $r_{0}<
r < \infty$, and then match two solutions at $r_{0}$ by
the match condition:
$$A_{j}(E,\lambda)\equiv \left. \displaystyle 
{f_{jE}(r,\lambda) \over g_{jE}(r,\lambda)}
\right|_{r=r_{0}-} = \left. \displaystyle 
{f_{jE}(r,\lambda) \over g_{jE}(r,\lambda)}
\right|_{r=r_{0}+} . \eqno (11) $$

\noindent
When $r_{0}$ is the zero point of $g_{jE}(r,\lambda)$, the 
match condition can be replaced by its inverse 
$g_{jE}(r,\lambda)/f_{jE}(r,\lambda)$ instead. The merit of 
using this match condition is that we need not care the 
normalization factor in the solutions.

The establishment of the Levinson theorem for the Dirac equation is 
similar to that for the Schr\"{o}dinger equation [20]. The main 
differences between them are that the ratio $f/g$ of two radial 
functions in the Dirac problem plays the role of the logarithmic 
derivative $R'/R$ in the Schr\"{o}dinger problem, and the
energy $E$ of bound states satisfies $|E|\leq M$ instead of $E\leq 0$. 
With the increment of the strength of the potential $V(r,\lambda)$,
the scattering state may turn into a bound state at the
energy $M$ or $-M$ and the bound state may also turn into 
a scattering state at the energies. 

The key point for the proof of the Levinson theorem is that 
the ratio $f_{jE}(r,\lambda)/g_{jE}(r,\lambda)$ is monotonic
with respect to the energy $E$, which is called the generalized
Sturm-Liouville theorem [21] and will be proved in Sec. II. 
Based on the generalized Sturm-Liouville theorem, in Sec.III
the number of bound states will be related with the 
variance of the ratio at $r_{0}-$ as the potential changes. 
In Sec.IV, we further prove this variance of the ratio
also determines the sum of the phase shifts at the energies
$\pm M$. In the course of proof, it can be seen evidently that as the 
potential changes, the phase shift at the energy $M$ jumps by 
$\pi$ while a scattering state of a positive energy becomes a 
bound state, and the phase shift at the energy $-M$ jumps by 
$-\pi$ while a bound state becomes a scattering state of 
a negative energy, or vice versa. The critical case, where 
the Dirac equation has a finite zero-momentum solution, will
be studied in Sec. IV in detail. A zero-momentum solution is 
called a half bound state if its wave function is finite but 
does not decay fast enough at infinity to be square integrable. 
Thus, the Levinson theorem relates the number $n_{j}$ of bound 
states with angular momentum $j$ to the sum of phase 
shifts $\eta_{j}(\pm M)$ with $j$ at the energies $\pm M$: 
$$ \eta_{j}(M)+\eta_{j}(-M)~~~~~~~~~~~~~~~~~~~~~~~~~~~~~~~~~~~~~~
~~~~~~~~~~~~~~~~~~~~~~~~~~~~~~~~~~~~~~~~~~~~~~~~~~~~~~~~~~~$$
$$~~~=\left\{\begin{array}{ll}
(n_{j}+1)\pi
&{\rm when~a~half~bound~state~occurs~at}~E=M ~~{\rm and}~~
j=3/2~{\rm or}~-1/2\\
(n_{j}+1)\pi
&{\rm when~a~half~bound~state~occurs~at}~E=-M~~{\rm and}~~
j=1/2~{\rm or}~-3/2\\
n_{j}\pi~&{\rm the~rest~cases} . \end{array} \right. 
\eqno (12) $$

\noindent
The problem that the potential has a tail at infinity will
be discussed in Sec.V.

\section{THE GENERALIZED STURM-LIOUVILLE THEOREM}

Suppose that $f,g$ and $f_{1},g_{1}$ are two solutions of Eq.(8)
with the energies $E$ and $E_{1}$, respectively. From Eq.(8) we have
$$\displaystyle {d \over dr}\left(f_{1}g-g_{1}f\right)
=-(E_{1}-E)\left(f_{1}f+g_{1}g\right), \eqno (13) $$

\noindent
From the boundary condition that both solutions vanish at the 
origin, we integrate Eq.(13) in the region $0\leq r\leq r_0$
and obtain
$$\left. \left(f_{1}g-g_{1}f\right)\right|_{r=r_{0}-}
=-(E_{1}-E)\displaystyle \int_{0}^{r_{0}}
\left(f_{1}f+g_{1}g\right)dr. $$

\noindent
Taking the limit as $E_{1}$ tends to $E$, we have
$$\left. \displaystyle \lim_{E_{1} \rightarrow E}
\displaystyle {f_{1}g-g_{1}f \over E_{1}-E}
\right|_{r=r_{0}-}
=\left\{g_{jE}(r_{0},\lambda)\right\}^{2} \displaystyle 
{\partial \over \partial E} A_{j}(E,\lambda)
=-\displaystyle \int_{0}^{r_{0}}\left\{f_{jE}^{2}(r,\lambda)
+g_{jE}^{2}(r,\lambda)\right\}dr< 0,  \eqno (14) $$

\noindent
where we denote the solution $f$ and $g$ by $f_{jE}(r_{0},\lambda)$
and $g_{jE}(r_{0},\lambda)$. Thus, when $|E|\geq M$ we have
$$A_{j}(E,\lambda)=A_{j}(M,\lambda)-c^{2}_{1}k^{2}+\ldots,~~~~~
{\rm when}~~E> M {\rm ~~and~~}E\sim M, $$
$$A_{j}(E,\lambda)=A_{j}(-M,\lambda)+c^{2}_{2}k^{2}+\ldots,~~~~~
{\rm when}~~E< -M {\rm ~~and~~}E\sim -M, \eqno (15)$$

\noindent
where $c^{2}_{1}$ and $c^{2}_{2}$ are non-negative numbers, and
the momentum $k$ is defined as follows:
$$k=\left(E^{2}-M^{2}\right)^{1/2}. \eqno (16) $$

Similarly, from the boundary condition that the radial functions 
$f_{jE}(r,\lambda)$ and $g_{jE}(r,\lambda)$ for $|E|\leq M$ 
tend to zero at infinity, we obtain by integrating Eq.(13) in 
the region $r_{0}\leq r < \infty$
$$\left\{g_{jE}(r_{0},\lambda)\right\}^{2}
\left. \displaystyle {\partial \over \partial E}
\left(\displaystyle {f_{jE}(r,\lambda) \over 
g_{jE}(r,\lambda)}\right)\right|_{r=r_{0}+}
= \displaystyle 
\int_{r_{0}}^{\infty}\left(f_{jE}^{2}(r,\lambda)+g_{jE}^{2}(r,\lambda)
\right)dr > 0.  \eqno (17) $$

\noindent
Thus, as the energy $E$ increases, the ratio 
$f_{jE}(r,\lambda)/g_{jE}(r,\lambda)$ at $r_{0}-$ 
$~\left(A_{j}(E,\lambda)\right)~$ decreases monotonically, but the ratio 
$f_{jE}(r,\lambda)/g_{jE}(r,\lambda)$ at $r_{0}+$ when $|E|\leq M$ 
increases monotonically. This is called the generalized 
Sturm-Liouville theorem [21].

\section{THE NUMBER OF BOUND STATES} 

Now, we solve Eq.(8) for the energy $|E|\leq M$. In the region
$0\leq r < r_{0}$, when $\lambda=0$, we have
$$f_{jE}(r,0)=e^{-i(j-1/2)\pi/2}\left\{(M+E)\pi \kappa r/2\right\}^{1/2}
J_{j-1/2}(i\kappa r), $$
$$g_{jE}(r,0)=e^{-i(j-3/2)\pi/2}\left\{(M-E)\pi \kappa r/2\right\}^{1/2}
J_{j+1/2}(i\kappa r), \eqno (18)  $$

\noindent
where $J_{m}(x)$ is the Bessel function, and 
$$\kappa=\left(M^{2}-E^{2}\right)^{1/2}. \eqno (19) $$

\noindent
The ratio at $r=r_{0}-$ when $\lambda=0$ is
$$A_{j}(E,0)=-i \left(\displaystyle {M+E \over M-E }\right)^{1/2}
 \displaystyle {J_{j-1/2}(i\kappa r_{0}) \over J_{j+1/2}(i\kappa r_{0}) }
~~~~~~~~~~~~~~~~~~~~~~~$$
$$~~~~~~~~~=\left\{\begin{array}{ll} - \displaystyle 
{2M(2j+1) \over \kappa^{2}r_{0} } \sim -\infty~~~~ 
&{\rm when}~~E\sim M \\[2mm]
- \displaystyle {2j+1 \over 2Mr_{0} }
 &{\rm when}~~E\sim -M . \end{array} \right.  \eqno (20) $$

In the region $r_{0}< r < \infty$, due to the cutoff 
potential we have $V(r)=0$ and
$$f_{jE}(r,\lambda)=e^{i(j+1/2)\pi/2}\left\{(M+E)\pi \kappa r/2\right\}^{1/2}
H^{(1)}_{j-1/2}(i\kappa r), $$
$$g_{jE}(r,\lambda)=e^{i(j+3/2)\pi/2}\left\{(M-E)\pi \kappa r/2\right\}^{1/2}
H^{(1)}_{j+1/2}(i\kappa r), \eqno (21) $$

\noindent
where $H^{(1)}_{m}(x)$ is the Hankel function of the first kind.
The ratio at $r=r_{0}+$ does not depend on $\lambda$ and is given
as follows:
$$\left. \displaystyle {f_{jE}(r,\lambda) \over g_{jE}(r,\lambda)} 
\right|_{r=r_{0}+}
=-i \left(\displaystyle {M+E \over M-E }\right)^{1/2} \displaystyle 
{H^{(1)}_{j-1/2}(i\kappa r_{0}) \over H^{(1)}_{j+1/2}(i\kappa r_{0}) } 
~~~~~~~~~~~~~~~~~~~~~~~~~~~~~~~~~~~~~~~~~~~~~~$$
$$~~~~~~~~~~~~~=\left\{\begin{array}{ll} \displaystyle 
{2Mr_{0} \over 2j-1 } &{\rm when}~~E\sim M ~~{\rm and}~~j\geq 3/2 \\
-2Mr_{0}\log(\kappa r_{0})\sim \infty~~~~ &{\rm when}~~E\sim M ~~{\rm and}~~j=1/
2 \\
\displaystyle {\kappa^{2}r_{0} \over 2M(2j-1) } \sim 0
 &{\rm when}~~E\sim -M , ~~{\rm and}~~j\geq 3/2 \\
\displaystyle {-\kappa^{2} r_{0}\log(\kappa r_{0}) \over 2M } \sim 0
 &{\rm when}~~E\sim -M , ~~{\rm and}~~j=1/2 .
\end{array} \right.  \eqno (22) $$

\noindent
It is evident from Eqs.(20) and (22) that as the energy $E$
increases from $-M$ to $M$, there is no overlap between two 
variant ranges of the ratio at two sides of $r_{0}$ when
$\lambda=0$ (no potential) except for $j=1/2$ where there is a 
half bound state at $E=M$. The half bound state will be discussed
in the next section.

As $\lambda$ increases from zero to one, the potential $V(r,\lambda)$
changes from zero to the given potential $V(r)$, and 
$A_{j}(E,\lambda)$ changes, too. If $A_{j}(M,\lambda)$ decreases
across the value $2Mr_{0}/(2j-1)$ as $\lambda$ increases, an 
overlap between the variant ranges of the ratios at two sides 
of $r_{0}$ appears. Since the ratio $A_{m}(E,\lambda)$ of two 
radial functions at $r_{0}-$ decreases monotonically as the energy 
$E$ increases, and the ratio at $r_{0}+$ increases monotonically, 
the overlap means that there must be one and only one energy
where the matching condition (11) is satisfied, namely a bound 
state appears. 

As $\lambda$ increases, $A_{j}(M,\lambda)$ may decreases to 
$-\infty$, jumps to $\infty$, and then decreases again across 
the value $2Mr_{0}/(2j-1)$, so that another bound state appears. 
Note that when $r_{0}$ is a zero point of the wave function 
$g_{jE}(r,\lambda)$, $A_{j}(E,\lambda)$ goes to infinity. It 
is not a singularity.

On the other hand, as $\lambda$ increases, if $A_{j}(-M,\lambda)$ 
decreases across zero, an overlap between the variant ranges 
of the ratios at two sides of $r_{0}$ disappears so that a 
bound state disappears. 

Therefore, each time $A_{j}(M,\lambda)$ decreases across the value 
$2Mr_{0}/(2j-1)$ as $\lambda$ increases, a new overlap between the 
variant ranges of the ratios at two sides of $r_{0}$ appears such 
that a scattering state of a positive energy becomes a bound 
state. On the other hand, each time $A_{j}(-M,\lambda)$ 
decreases across zero, an overlap between the variant ranges 
of the ratio at two sides of $r_{0}$ disappears such that a 
bound state becomes a scattering state of a negative energy. 
Conversely, each time $A_{j}(M,\lambda)$ increases
across the value $2Mr_{0}/(2j-1)$, an overlap between the 
variant ranges disappears such that a bound state becomes a
scattering state of a positive energy, and each time $A_{j}(-M,\lambda)$ 
increases across zero, a new overlap between the variant ranges 
appears such that a scattering state of a negative energy becomes 
a bound state. 

Now, the number $n_{j}$ of bound states with the angular momentum $j$
is equal to the sum (or subtraction) of four times as $\lambda$ 
increases from zero to one: the times that $A_{j}(M,\lambda)$ 
decreases across the value $2Mr_{0}/(2j-1)$, minus the times 
that $A_{j}(M,\lambda)$ increases across the value $2Mr_{0}/(2j-1)$, 
minus the times that $A_{j}(-M,\lambda)$ decreases across zero, 
plus the times that $A_{j}(-M,\lambda)$ increases across zero.

When $j=1/2$, the value $2Mr_{0}/(2j-1)$ becomes
infinity. We may check the times that $A_{j}(M,\lambda)^{-1}$ 
increases (or decreases) across zero to replace the times that 
$A_{j}(M,\lambda)$ decreases (or increases) across infinity.

\section{THE RELATIVISTIC LEVINSON THEOREM}

We turn to discuss the phase shifts of the scattering states.
Solving Eq.(8) in the region $r_{0}<r<\infty$ for the energy $|E|>M$,
we have
$$f_{jE}(r,\lambda)=B(E) \left(\displaystyle {\pi kr \over 2 } \right)^{1/2}
\left\{ \cos \eta_{j}(E,\lambda)J_{j-1/2}(kr)
-\sin \eta_{j}(E,\lambda) N_{j-1/2}(kr)\right\}, $$
$$g_{jE}(r,\lambda)=\left(\displaystyle {\pi kr \over 2 } \right)^{1/2}
\left\{\cos \eta_{j}(E,\lambda)J_{j+1/2}(kr)
-\sin \eta_{j}(E,\lambda) N_{j+1/2}(kr)\right\}, 
\eqno (23) $$

\noindent
where $N_{m}(x)$ denotes the Neumann function, the momentum
$k$ is given in Eq.(16), and $B(E)$ is defined as
$$B(E)=\left\{\begin{array}{ll}
\left(\displaystyle {E+M \over E-M}\right)^{1/2} &{\rm when}~~
E>M \\
-\left(\displaystyle {|E|-M \over |E|+M}\right)^{1/2} &{\rm when}~~
E<-M . \end{array} \right. \eqno (24) $$

\noindent
The asymptotic form of the solution (23) at $r\longrightarrow \infty$
is
$$f_{jE}(r,\lambda)\sim B(E) \cos \left(kr-j\pi/2+\eta_{j}(E,\lambda)
\right), $$
$$g_{jE}(r,\lambda)\sim  \sin \left(kr-j\pi/2+\eta_{j}(E,\lambda)
\right). \eqno (25) $$

\noindent
Substituting Eq.(23) into the match condition (11), we obtain
the formula for the phase shift $\eta_{j}(E,\lambda)$:
$$\tan \eta_{j}(E,\lambda)=\displaystyle {J_{j+1/2}(kr_{0})
\over N_{j+1/2}(kr_{0})}~\cdot~\displaystyle 
{A_{j}(E,\lambda)-B(E)J_{j-1/2}(kr_{0})/J_{j+1/2}(kr_{0})\over
A_{j}(E,\lambda)-B(E)N_{j-1/2}(kr_{0})/N_{j+1/2}(kr_{0}) } $$
$$~~~~~~~~~~~~~~~~~~~~~~=\displaystyle {J_{j-1/2}(kr_{0})
\over N_{j-1/2}(kr_{0})}~\cdot~\displaystyle 
{\left\{A_{j}(E,\lambda)\right\}^{-1}-B(E)^{-1}J_{j+1/2}(kr_{0})/
J_{j-1/2}(kr_{0})
\over \left\{A_{j}(E,\lambda)\right\}^{-1}-B(E)^{-1}N_{j+1/2}(kr_{0})/
N_{j-1/2}(kr_{0}) } .\eqno (26) $$

\noindent
The phase shift $\eta_{j}(E,\lambda)$ is determined up to
a multiple of $\pi$ due to the period of the tangent
function. We use the convention that the phase shifts
for the free particles ($V(r)=0$) are vanishing:
$$\eta_{j}(E,0)=0. \eqno (27) $$

\noindent
Under this convention, the phase shifts $\eta_{j}(E)$ are determined 
completely as $\lambda$ increases from zero to one:
$$\eta_{j}(E)\equiv \eta_{j}(E,1) \eqno (28) $$

The phase shifts $\eta_{j}(\pm M,\lambda)$ are the limits of 
the phase shifts $\eta_{j}(E,\lambda)$ as $E$ tends to $\pm M$.
At the sufficiently small $k$, $k\ll 1/r_{0}$, we have
$$\tan \eta_{j}(E,\lambda)~~~~~~~~~~~~~~~~~~~~~~~~~~~~~~~~~~~~~~~~~~~
~~~~~~~~~~~~~~~~~~~~~~~~~~~~~~~~~~~~~~~~~~~~~~~~~~~~~~~~~~~~~~~~~~~~~$$
$$~~~~\sim \left\{\begin{array}{ll}
-\displaystyle {\pi (kr_{0}/2)^{2j+1} \over (j+1/2)!(j-1/2)!}
\cdot \displaystyle {
A_{j}(M,\lambda)-2M(2j+1)/(k^{2}r_{0}) \over
A_{j}(M,\lambda)-c^{2}_{1}k^{2}-\displaystyle {2Mr_{0}\over 2j-1}
\left(1+\displaystyle {(kr_{0})^{2} \over (2j-1)(2j-3)} \right) }
&{\rm when}~~j>3/2 \\[2mm]
-\displaystyle {\pi \over 2}~ \left(\displaystyle {kr_{0} \over 2}\right)^{4}
~\cdot~\displaystyle {
A_{j}(M,\lambda)-8M/(k^{2}r_{0}) \over
A_{j}(M,\lambda)-c^{2}_{1}k^{2}-Mr_{0}
\left(1-\displaystyle {\left(kr_{0}\right)^{2} \over 2 } 
\log(kr_{0}) \right) }
&{\rm when}~~j=3/2 \\[2mm]
\displaystyle {\pi \over 2 \log(kr_{0})}
~\cdot~\displaystyle {
\left\{A_{j}(M,\lambda)\right\}^{-1}+c^{2}_{1}k^{2}-k^{2}r_{0}/(4M) \over
\left\{A_{j}(M,\lambda)\right\}^{-1}+c^{2}_{1}k^{2}+\left\{ 
2Mr_{0}\log(kr_{0})\right\}^{-1} }
&{\rm when}~~j=1/2 , \end{array} \right.  \eqno (29) $$

\noindent
for $E>M$, and 
$$\tan \eta_{j}(E,\lambda)
\sim \left\{\begin{array}{ll}
-\displaystyle {\pi (kr_{0}/2)^{2j+1} \over (j+1/2)!(j-1/2)!}
~\cdot~\displaystyle {
A_{j}(-M,\lambda)+(2j+1)/(2Mr_{0}) \over
A_{j}(-M,\lambda)+c^{2}_{2}k^{2}+\displaystyle {k^{2}r_{0}\over 2M(2j-1)} }
&{\rm when}~~j\geq 3/2 \\[2mm]
-\pi \left( \displaystyle {kr_{0} \over 2}\right)^{2}
~\cdot~\displaystyle {
A_{j}(-M,\lambda)+1/(Mr_{0}) \over
A_{j}(-M,\lambda)+c^{2}_{2}k^{2}-k^{2}r_{0}\log(kr_{0})/(2M)}
&{\rm when}~~j=1/2 , \end{array} \right. \eqno (30) $$

\noindent
for $E<-M$. The asymptotic forms (15) have been used in driving 
Eqs.(29) and (30). In addition to the leading terms, we include
in Eqs.(29) and (30) some next leading terms, which are useful
only for the critical case where the leading terms are canceled 
to each other.

Firstly, from Eqs.(29) and (30) we see that $\tan \eta_{j}(E,\lambda)$
tends to zero as $E$ goes to $\pm M$, namely, $\eta_{j}(\pm M, \lambda)$
are always equal to the multiple of $\pi$. In other words, if the
phase shift $\eta_{j}(E,\lambda)$ for a sufficiently small $k$
is expressed as a positive or negative acute angle plus $n\pi$,
its limit $\eta_{j}(M,\lambda)$ (or $\eta_{j}(-M,\lambda)$)
is equal to $n\pi$. It means that $\eta_{j}(M,\lambda)$ (or
$\eta_{j}(-M,\lambda)$) changes discontinuously when
$\eta_{j}(E,\lambda)$ changes through the value $(n+1/2)\pi$,where $n$ is an
integer.

Secondly, from Eq.(26) we have
$$\left. \displaystyle {\partial \eta_{j}(E,\lambda) \over \partial
A_{j}(E,\lambda) }\right|_{E} =-\left(\displaystyle {E+M \over E-M}
\right)^{1/2} \displaystyle {2\left\{\cos \eta_{j}(E,\lambda)\right\}^{2}
\over \pi kr_{0} \left\{N_{j+1/2}(kr_{0})A_{j}(E,\lambda)
-B(E)N_{j-1/2}(kr_{0}) \right\}^{2} }\leq 0,~~~~~E>M, $$
$$\left. \displaystyle {\partial \eta_{j}(E,\lambda) \over \partial
A_{j}(E,\lambda)} \right|_{E} =\left(\displaystyle {|E|-M \over |E|+M}
\right)^{1/2} \displaystyle {2\left\{\cos \eta_{j}(E,\lambda)\right\}^{2}
\over \pi kr_{0} \left\{N_{j+1/2}(kr_{0})A_{j}(E,\lambda)
-B(E)N_{j-1/2}(kr_{0}) \right\}^{2} }\geq 0,~~~~~E<-M.
\eqno (31) $$

\noindent
Namely, as the ratio $A_{j}(E,\lambda)$ decreases, the phase shift
$\eta_{j}(E, \lambda)$ for $E>M$ increases monotonically,
but $\eta_{j}(E, \lambda)$ for $E<-M$ decreases monotonically.
In terms of the monotonic properties we are able to determine
the jump of the phase shifts $\eta_{j}(\pm M,\lambda)$. 

We first consider the scattering states of a positive energy 
with a sufficiently small momentum $k$. As $A_{j}(E,\lambda)$ 
decreases, if $\tan \eta_{j}(E,\lambda)$ changes sign from 
positive to negative, the phase shift $\eta_{j}(M,\lambda)$ 
jumps by $\pi$. Note that in this case if $\tan \eta_{j}(E,\lambda)$ 
changes sign from negative to positive, the phase shift 
$\eta_{j}(M,\lambda)$ keeps invariant. Conversely, as 
$A_{j}(E,\lambda)$ increases, if $\tan \eta_{j}(E,\lambda)$ 
changes sign from negative to positive, the phase shift 
$\eta_{j}(M,\lambda)$ jumps by $-\pi$. Therefore, as $\lambda$ 
increases from zero to one, each time the $A_{j}(M,\lambda)$ 
decreases from near and larger than the value $2Mr_{0}/(2j-1)$ 
to smaller than that value, the denominator in Eq.(29) changes 
sign from positive to negative and the rest factor keeps positive, 
so that the phase shift $\eta_{j}(M,\lambda)$ jumps by $\pi$. It 
has been shown in the previous section that each time the 
$A_{j}(M,\lambda)$ decreases across the value $2Mr_{0}/(2j-1)$, 
a scattering state of a positive energy becomes a bound state. 
Conversely, each time the $A_{j}(M,\lambda)$ increases across 
that value, the phase shift $\eta_{j}(M,\lambda)$ jumps by $-\pi$, 
and a bound state becomes a scattering state of a positive energy. 

Then, we consider the scattering states of a negative energy with a 
sufficiently small $k$. As $A_{j}(E,\lambda)$ decreases, if 
$\tan \eta_{j}(E,\lambda)$ changes sign from negative to positive,
the phase shift $\eta_{j}(-M,\lambda)$ jumps by $-\pi$.
However, in this case if $\tan \eta_{j}(E,\lambda)$ changes sign 
from positive to negative, the phase shift $\eta_{j}(-M,\lambda)$
keeps invariant. Conversely, as $A_{j}(E,\lambda)$ increases, if 
$\tan \eta_{j}(E,\lambda)$ changes sign from positive to negative,
the phase shift $\eta_{j}(-M,\lambda)$ jumps by $\pi$. Therefore, 
as $\lambda$ increases from zero to one, each time the 
$A_{j}(-M,\lambda)$ decreases from a small and positive
number to a negative one, the denominator in Eq.(29) changes sign 
from positive to negative and the rest factor keeps negative, so 
that the phase shift $\eta_{j}(-M,\lambda)$ jumps by $-\pi$. In 
the previous section it is shown that each time the $A_{j}(-M,\lambda)$ 
decreases across zero, a bound state becomes a scattering state of 
a negative energy. Conversely, each time the $A_{j}(-M,\lambda)$ 
increases across zero, the phase shift $\eta_{j}(-M,\lambda)$ 
jumps by $\pi$, and a scattering state of a negative energy
becomes a bound state. Therefore, we obtain the Levinson theorem 
for the Dirac equation in two dimensions for non-critical cases:
$$\eta_{j}(M)+\eta_{j}(-M)=n_{j} \pi. \eqno (32) $$

\noindent
It is obvious that the Levinson theorem (32) holds for
both positive and negative $j$ in the non-critical cases.

For the case of $j=1/2$ and $E\sim M$, where the value 
$2Mr_{0}/(2j-1)$ is infinity. Since 
$\left\{A_{j}(E,\lambda)\right\}^{-1}$ increases as $A_{j}(E,\lambda)$ 
decreases, we can study the variance of 
$\left\{A_{j}(E,\lambda)\right\}^{-1}$ in this case instead.
For the energy $E>M$ where the momentum $k$ is sufficiently small, 
when $\left\{A_{j}(M,\lambda)\right\}^{-1}$ increases from negative 
to positive as $\lambda$ increases, both the numerator and denominator in 
Eq.(29) change signs, but not simultaneously. The numerator changes 
sign first, and then, the denominator changes. The front factor in
Eq.(29) is negative so that $\tan \eta_{j}(E,\lambda)$ firstly
changes from negative to positive when the numerator changes sign,
and then changes from positive to negative when the denominator
changes sign. It is in the second step that the phase shift
$\eta_{j}(M,\lambda)$ jumps by $\pi$. Similarly, each time 
$\left\{A_{j}(M,\lambda)\right\}^{-1}$ decreases across zero 
as $\lambda$ increases, $\eta_{j}(M,\lambda)$ jumps by $-\pi$. 

For $\lambda=0$ and $j=1/2$, the numerator in Eq.(29) is equal 
to zero, and the phase shift $\eta_{j}(M,0)$ is defined to be 
zero. For this case there is a half bound state at $E=M$ (see 
Eq.(33)). If $\left\{A_{j}(M,\lambda)\right\}^{-1}$ increases 
($A_{j}(M,\lambda)$ decreases) as $\lambda$ increases from zero, 
the front factor in Eq.(29) is negative, the numerator becomes positive 
first, and then, the denominator changes sign from negative
to positive, such that the phase shift $\eta_{j}(M,\lambda)$
jumps by $\pi$ and simultaneously the half bound state becomes
a bound state with $E<M$.

Now, we turn to study the critical cases. Firstly, we study the 
critical case for $E\sim M$, where the ratio $A_{j}(M,1)$ is equal 
to the value $2Mr_{0}/(2j-1)$. It is easy to obtain the following
solution of $E=M$ in the region $r_{0}<r<\infty$ , satisfying the 
radial equations (8) and the match condition (11) at $r_{0}$:
$$f_{jM}(r,1)=2Mr^{-j+1},~~~~~g_{jM}(r,1)=(2j-1)r^{-j}. \eqno (33) $$

\noindent
It is a bound state when $j> 3/2$, but called a half bound state
when $j=3/2$ or $j=1/2$. A half bound state is not a bound
state, because its wave function is finite but not square integrable. 

For definiteness, we assume that in the critical case, as 
$\lambda$ increases from a number near and less than one and
finally reaches one, $A_{j}(M,\lambda)$ 
{\it decreases} and finally reaches, but not across, the value 
$2Mr_{0}/(2j-1)$. In this case, when $\lambda=1$ a new bound 
state of $E=M$ appears for $j>3/2$, but does not appear
for $j=3/2$ or $j=1/2$. We should check whether or not 
the phase shift $\eta_{j}(M,1)$ increases by an additional 
$\pi$ as $\lambda$ increases and reaches one.

It is evident from the next leading terms in the 
denominator of Eq.(29) that the denominator for $j\geq 3/2$ 
has changed sign from positive to negative as $A_{j}(M,\lambda)$ 
decreases and finally reaches the value $2Mr_{0}/(2j-1)$, 
namely, the phase shift $\eta_{j}(M,\lambda)$ jumps by an
additional $\pi$ at $\lambda=1$. Simultaneously, a new bound 
state of $E=M$ appears for $j>3/2$, but only a half bound state
appears for $j=3/2$, so that the Levinson theorem (32) holds 
for the critical case with $j>3/2$, but it has to be modified
for the critical case with $j=3/2$:
$$\eta_{j}(M)+\eta_{j}(-M)=(n_{j}+1)\pi,~~~~~{\rm when~a~half~
bound~state~occurs~at}~~E=M~{\rm and}~j=3/2. \eqno (34) $$

For $j=1/2$ the next leading term with log$(kr_{0})$ in the 
denominator of Eq.(29) dominates so that the denominator keeps 
negative (does not change sign!) as 
$\left\{A_{j}(M,\lambda)\right\}^{-1}$ increases and finally 
reaches zero, namely, the phase shift $\eta_{j}(M,\lambda)$ 
does not jump, no matter whether the rest part in Eq.(28) keeps
positive or has changed to negative. Simultaneously, only a new 
half bound state of $E=M$ for $j=1/2$ appears, so that the 
Levinson theorem (32) holds for the critical case with $j=1/2$. 

This conclusion holds for the critical case where $A_{j}(M,\lambda)$ 
{\it increases} and finally reaches, but not across, the value 
$2Mr_{0}/(2j-1)$.

Secondly, we study the critical case for $E=-M$, where the ratio 
$A_{j}(-M,1)$ is equal to zero. It is easy to obtain the following
solution of $E=-M$ in the region $r_{0}<r<\infty$, satisfying the 
radial equations (8) and the match condition (11) at $r_{0}$:
$$f_{jM}(r,\lambda)=0,~~~~~g_{jM}(r,\lambda)=r^{-j}. \eqno (35) $$

\noindent
It is a bound state when $j\geq 3/2$, but a half bound state
when $j=1/2$. 

For definiteness, we again assume that in the critical case, as 
$\lambda$ increases from a number near and less than one and finally
reaches one, $A_{j}(-M,\lambda)$ {\it decreases} and finally reaches zero, 
so that when $\lambda=1$ the energy of a bound state decreases
to $E=-M$ for $j\geq 3/2$, but a bound state becomes a half bound 
state for $j=1/2$. We should check whether or not the phase shift 
$\eta_{j}(-M,1)$ decreases by $\pi$ as $\lambda$ increases and reaches one.

For the energy $E< -M$ where the momentum $k$ is sufficiently small, 
one can see from the next leading terms in the denominator of Eq.(30) 
that the denominator does not change sign as $A_{j}(-M,\lambda)$ 
decreases and finally reaches zero, namely, the phase shift 
$\eta_{j}(-M,\lambda)$ does not jump by an additional $-\pi$ at 
$\lambda=1$. Simultaneously, the energy of a bound state decreases 
to $E=-M$ for $j\geq 3/2$, but a bound state becomes a half bound state
for $j=1/2$, so that the Levinson theorem (32) holds 
for the critical case with $j\geq 3/2$, but it has to be modified
for the critical case with $j=1/2$:
$$\eta_{j}(M)+\eta_{j}(-M)=(n_{j}+1)\pi,~~~~~{\rm when~a~half~
bound~state~occurs~at}~~E=-M~{\rm and}~j=1/2. \eqno (36) $$

Combining Eqs.(32), (34), (36) and their corresponding forms for the
negative $j$, we obtain the relativistic Levinson theorem (12) in 
two dimensions.

\section{DISCUSSIONS}

Now, we discuss the general case where the potential $V(r)$
has a tail at $r\geq r_{0}$. Let $r_{0}$ is so large that
only the leading term in $V(r)$ is concerned:
$$V(r) \sim br^{-n},~~~~~r\geq r_{0}. \eqno (37) $$

\noindent
where $b$ is a non-vanishing constant and $n$ is a positive constant,
not necessary to be an integer. Substituting it into 
Eq.(8) and changing the variable $r$ to $\xi$:
$$\xi=\left\{\begin{array}{ll}
kr=r\sqrt{E^{2}-M^{2}},~~~~&{\rm when}~~|E|>M \\
\kappa r=r\sqrt{M^{2}-E^{2}},~~~~&{\rm when}~~|E|\leq M, 
\end{array} \right. \eqno (38) $$

\noindent
we obtain the radial equations in the region $r_{0}\leq r < \infty$:
$$\displaystyle {d \over d\xi}g_{jE}(\xi)+\displaystyle 
{j \over \xi} g_{jE}(\xi) 
=\left(\displaystyle {E \over |E|}
\sqrt{\displaystyle {E-M \over E+M} }-
\displaystyle {b \over \xi^{n}} k^{n-1}\right)f_{jE}(\xi) , $$
$$-\displaystyle {d \over dr}f_{jE}(\xi)+\displaystyle 
{j \over r} f_{jE}(\xi) 
=\left(\displaystyle {E \over |E|}\sqrt{\displaystyle {E+M \over E-M} }-
\displaystyle {b \over \xi^{n}} k^{n-1}\right)g_{jE}(\xi) ,
\eqno (39) $$

\noindent
for $|E|>M$, and 
$$\displaystyle {d \over d\xi}g_{jE}(\xi)+\displaystyle 
{j \over \xi} g_{jE}(\xi) 
=\left(- \sqrt{\displaystyle {M-E \over M+E} }-
\displaystyle {b \over \xi^{n}} \kappa^{n-1}\right)f_{jE}(\xi) , $$
$$-\displaystyle {d \over dr}f_{jE}(\xi)+\displaystyle 
{j \over r} f_{jE}(\xi) 
=\left(\sqrt{\displaystyle {M+E \over M-E} }-
\displaystyle {b \over \xi^{n}} \kappa^{n-1}\right)g_{jE}(\xi) ,
\eqno (40) $$

\noindent
for $|E|\leq M$. As far as the Levinson theorem is concerned, 
we are only interested in the solutions with the sufficiently
small $k$ and $\kappa$. If $n\geq 3$, in comparison with the
first term on the right hand side of Eq.(39) or Eq.(40), the 
potential term with a factor $k^{n-1}$ (or $\kappa^{n-1}$) is 
too small to affect the phase shift at the sufficiently small $k$ 
and the variant range of the ratio $f_{jE}(r,\lambda)/g_{jE}(r,\lambda)$
at $r_{0}+$. Therefore, the proof given in the previous sections
is effective for those potentials with a tail so that the
Levinson theorem (12) holds. 

When $n=2$ and $b\neq 0$, we will only keep the leading terms
for the small parameter $k$ (or $\kappa$) in solving Eq.(39)
(or Eq.(40)). Firstly, we calculate the solutions with the 
energy $E\sim M$. Let
$$\alpha=\left(j^{2}-j+2Mb+1/4\right)^{1/2}\neq j-1/2. \eqno (41) $$

\noindent
If $\alpha^{2}<0$, there is an infinite number of bound states. 
We will not discuss this case as well as the case with $\alpha=0$
here. When $\alpha^{2}>0$, we take $\alpha>0$ for convenience. 
Some formulas given in the previous sections will be changed.

When $E\leq M$ we have
$$f_{jE}(r,\lambda)=e^{i(\alpha+1)\pi/2}2M\left(\pi \kappa r /2\right)^{1/2}
H^{(1)}_{\alpha}(i\kappa r),~~~~~~~~~~~~~~~~~~~~~~~~~~~~~
~~~~~~~~~~~~~ $$
$$g_{jE}(r,\lambda)=e^{i(\alpha+1)\pi/2}\kappa \left(\pi \kappa r /2
\right)^{1/2}
\left\{-\displaystyle {d \over d(\kappa r)}
H^{(1)}_{\alpha}(i\kappa r)+\displaystyle {j-1/2 \over \kappa r}
H^{(1)}_{\alpha}(i\kappa r)\right\}. \eqno (42)  $$

\noindent
Hence, the ratio at $r=r_{0}+$ for $E=M$ is
$$\left. \displaystyle {f_{jE}(r,\lambda) \over g_{jE}(r,\lambda) } 
\right|_{r=r_{0}+}
=\displaystyle {2Mr_{0} \over j+\alpha-1/2},~~~~~E=M. \eqno (43) $$

When $E> M$ we have
$$f_{jE}(r,\lambda)=2M\left(\pi k r /2\right)^{1/2}
\left\{ \cos \delta_{\alpha}(E,\lambda)J_{\alpha}(k r)
-\sin \delta_{\alpha}(E,\lambda)N_{\alpha}(k r) \right\}, $$
$$g_{jE}(r,\lambda)=k\left(\pi k r /2\right)^{1/2}
\left\{ \cos \delta_{\alpha}(E,\lambda)\left(-\displaystyle {d \over d(kr)}
J_{\alpha}(k r) +\displaystyle {j-1/2 \over kr} J_{\alpha}(kr) \right) 
\right.~~~~$$
$$\left.~~~~~~~~~~~-\sin \delta_{\alpha}(E,\lambda)
\left(-\displaystyle {d \over d(kr)}
N_{\alpha}(k r) +\displaystyle {j-1/2 \over kr} N_{\alpha}(kr) \right) 
\right\}, \eqno (44) $$

\noindent
When $kr$ tends to infinity, the asymptotic form
of the solution is:
$$f_{jE}(r,\lambda)\sim 2M\cos\left(kr-\alpha \pi/2-\pi/4
+\delta_{\alpha}(E,\lambda)\right), $$
$$g_{jE}(r,\lambda)\sim k \sin\left(kr-\alpha \pi/2-\pi/4
+\delta_{\alpha}(E,\lambda)\right). $$

\noindent
In comparison with the solution (25) we obtain the phase 
shift $\eta_{j}(E,\lambda)$ for $E>M$:
$$\eta_{j}(E,\lambda)=\delta_{\alpha}(E,\lambda)+
\left(j-\alpha -1/2\right)\pi/2,~~~~~E>M . \eqno (45) $$

\noindent
From the match condition (11), for the sufficiently small $k$ 
we obtain
$$\tan \delta_{\alpha}(E,\lambda)\sim \displaystyle 
{-\pi (kr_{0}/2)^{2\alpha} \over \Gamma(\alpha+1)\Gamma(\alpha) }
\left(\displaystyle {j-\alpha-1/2 \over j+\alpha-1/2 }\right)
\displaystyle {A_{j}(M,\lambda)-2Mr_{0}/(j-\alpha-1/2) \over
A_{j}(M,\lambda)-2Mr_{0}/(j+\alpha-1/2) }. \eqno (46) $$

\noindent
Therefore, as $\lambda$ increases from zero to one, each time 
the $A_{j}(M,\lambda)$ decreases from near and larger 
than the value $2Mr_{0}/(j+\alpha-1/2)$ to smaller than that value, 
the denominator in Eq.(46) changes sign from positive to negative 
and the rest factor keeps positive, so that $\delta_{\alpha}(M,\lambda)$ 
jumps by $\pi$. Simultaneously, from Eq.(43) a new overlap between the 
variant ranges of the ratio at two sides of $r_{0}$ appears
such that a scattering state of a positive energy becomes a bound 
state. Conversely, each time the $A_{j}(M,\lambda)$ 
increases across that value, $\delta_{\alpha}(M,\lambda)$ 
jumps by $-\pi$, and a bound state becomes a scattering state of 
a positive energy. 

Secondly, we calculate the solutions with the energy $E\sim -M$.
Let
$$\beta=\left(j^{2}+j-2Mb+1/4\right)^{1/2}\neq j+1/2. \eqno (47) $$

\noindent
Similarly, we only discuss the cases with $\beta^{2}>0$, and
take $\beta>0$. 

When $E\geq -M$ we have
$$f_{jE}(r,\lambda)=-e^{i(\beta+1)\pi/2}\kappa \left(\pi \kappa r /2
\right)^{1/2} \left\{\displaystyle {d \over d(\kappa r)}
H^{(1)}_{\beta}(i\kappa r)+\displaystyle {j+1/2 \over \kappa r}
H^{(1)}_{\beta}(i\kappa r)\right\}, $$
$$g_{jE}(r,\lambda)=e^{i(\beta+1)\pi/2}2M\left(\pi \kappa r /2\right)^{1/2}
H^{(1)}_{\beta}(i\kappa r).~~~~~~~~~~~~~~~~~~~~~~~~~~~~~~~~
~~~~~~~~  \eqno (48)  $$

\noindent
Hence, the ratio at $r=r_{0}+$ for $E=-M$ is
$$\left. \displaystyle {f_{jE}(r,\lambda) \over g_{jE}(r,\lambda) } 
\right|_{r=r_{0}+}
=- \displaystyle {j-\beta +1/2 \over 2Mr_{0}},~~~~~E=-M. \eqno (49) $$

When $E< -M$ we have
$$f_{jE}(r,\lambda)=-k\left(\pi k r /2\right)^{1/2}
\left\{ \cos \delta_{\beta}(E,\lambda)\left(\displaystyle {d \over d(kr)}
J_{\beta}(k r) +\displaystyle {j+1/2 \over kr} J_{\beta}(kr) \right)
\right.~~~~$$
$$\left.-\sin \delta_{\beta}(E,\lambda)\left(\displaystyle {d \over d(kr)}
N_{\beta}(k r) +\displaystyle {j+1/2 \over kr} N_{\beta}(kr) \right) 
\right\}, $$
$$g_{jE}(r,\lambda)=2M\left(\pi k r /2\right)^{1/2}
\left\{ \cos \delta_{\beta}(E,\lambda)J_{\beta}(k r)
-\sin \delta_{\beta}(E,\lambda)N_{\beta}(k r) \right\},  \eqno (50) $$

\noindent
When $kr$ tends to infinity, the asymptotic form
for the solution is:
$$f_{jE}(r,\lambda)\sim k\sin\left(kr-\beta \pi/2-\pi/4
+\delta_{\beta}(E,\lambda)\right), $$
$$g_{jE}(r,\lambda)\sim 2M \cos\left(kr-\beta \pi/2-\pi/4
+\delta_{\beta}(E,\lambda)\right). $$

\noindent
In comparison with the solution (25) we obtain the phase 
shift $\eta_{j}(E,\lambda)$ for $E<-M$:
$$\eta_{j}(E,\lambda)=\delta_{\beta}(E,\lambda)+
\left(j-\beta +1/2\right)\pi/2,~~~~~E<-M . \eqno (51) $$

\noindent
From the match condition (11), for the sufficiently small $k$
we obtain
$$\tan \delta_{\alpha}(E,\lambda)\sim \displaystyle 
{-\pi (kr_{0}/2)^{2\beta} \over \Gamma(\beta+1)\Gamma(\beta) }~\cdot~
\displaystyle {A_{j}(-M,\lambda)+(j+\beta+1/2)/(2Mr_{0}) \over
A_{j}(-M,\lambda)+(j-\beta+1/2)/(2Mr_{0}) }. \eqno (52) $$

\noindent
Therefore, as $\lambda$ increases from zero to one, each time 
the $A_{j}(-M,\lambda)$ decreases from near and larger 
than the value $-(j-\beta+1/2)/(2Mr_{0})$ to smaller than that value, 
the denominator in Eq.(52) changes sign from positive to negative 
and the rest factor keeps negative, so that $\delta_{\beta}(-M,\lambda)$ 
jumps by $-\pi$. Simultaneously, from Eq.(49) an overlap between the 
variant ranges of the ratio at two sides of $r_{0}$ disappears
such that a bound state becomes a scattering state of a negative 
energy. Conversely, each time the $A_{j}(-M,\lambda)$ 
increases across that value, $\delta_{\beta}(-M,\lambda)$ 
jumps by $\pi$, and a scattering state of a negative energy
becomes a bound state. 

In summary, we obtain the modified relativistic Levinson
theorem for non-critical cases when the potential has a tail (37)
with $n=2$ at infinity:
$$\eta_{j}(M)+\eta_{j}(-M)=n_{j}\pi +\left(2j-\alpha-\beta\right)\pi/2 .
\eqno (53) $$

We will not discuss the critical cases in detail. In fact,
the modified relativistic Levinson theorem (53) holds for 
the critical cases of $\alpha>1$ and $\beta>1$. When $0<\alpha<1$ 
or $0<\beta < 1$, $\delta_{\alpha}(M,1)$ or $\delta_{\beta}(-M,1)$
in the critical case will not be multiple of $\pi$, respectively, 
so that Eq.(53) is violated for those critical cases.

Furthermore, for the potential (37) with a tail at infinity,
when $n>2$, even if it contains a logarithm factor, for any 
arbitrarily small positive $\epsilon$, one can always find a 
sufficiently large $r_{0}$ such that $|V(r)|<\epsilon/r^{2}$ 
in the region $r_{0}<r<\infty$. Thus, from Eqs.(41) and (47)
we have for the sufficiently small $\epsilon$
$$\alpha=\left(j^{2}-j \pm 2M\epsilon +1/4\right)^{1/2}
\sim j-1/2, $$
$$\beta=\left(j^{2}+j \mp 2M\epsilon +1/4\right)^{1/2}
\sim j+1/2. $$

\noindent
Hence, equation (53) coincides with Eq.(32). In this case the
Levinson theorem (32) holds for the non-critical case.

\vspace{4mm}
{\bf ACKNOWLEDGMENTS}. This work was supported by the National
Natural Science Foundation of China and Grant No. LWTZ-1298 of
the Chinese Academy of Sciences.

\end{document}